\documentstyle[11pt,paspconf]{article}
\input{epsf}
\begin{document}

\title{Protoneutron stars and constraints on maximum and minimum mass
of neutron stars}

\author{Gondek D., Haensel P., Zdunik J. L.}
\affil{ Nicolaus Copernicus Astronomical Center, Polish
Academy of Sciences, Bartycka 18, 00-716 Warszawa, Poland}

\begin{abstract}
Constraints on minimum and maximum mass of ordinary neutron stars are
imposed by the consideration of their early evolution (protoneutron star
stage). Calculations are performed for a realistic standard model of
hot, dense matter (Lattimer \& Swesty 1991) valid for both
supranuclear and subnuclear densities.  Various assumptions concerning
protoneutron star interior (large trapped lepton number, no trapped
lepton number, isentropic, isothermal) are taken into account.
\end{abstract}



\section{Introduction}
The heavy-element core of massive, evolved stars ($M_{star} > 8\ M_\odot$) is
believed to collapse either directly to a black hole (BH) or to a metastable
protoneutron star (PNS).  The newborn neutron star does not resemble
the cold ordinary neutron star (NS).  It is a very hot object, with
temperature $T> 10\ \rm MeV$ and radius significantly larger than
that of a cold neutron star with the same number of baryons.  In
contrast to ordinary cold NS it is rich in leptons: electrons and
trapped degenerate or non-degenerate neutrinos.  A few seconds after
birth $t \simeq 2-4 \ {\rm s}$ the matter in the core of a hot neutron star has
almost constant lepton fraction ($Y_l=0.3-0.4$) and entropy per baryon
($s=1-2$ in unit of the Boltzmann constant $k_B$) (Burrows \& Lattimer
1986). PNS evolves either to BH or to stable NS depending on
its total number of baryons. A hot neutron star transforms into NS on 
a timescale 20-30 s.

Constraints on maximum mass of neutron stars imposed by composition
and equation of state (EOS) of hot dense stellar interior were studied
by numerous authors (e.g. Takatsuka 1995, Bombaci et. al 1995, Bombaci
1996, Prakash et al. 1997). For low densities of PNS they took EOS of
cold matter (at T=0).  In contrast to them we use a unified hot dense
matter model, which holds for both supranuclear and subnuclear
densities. We find the position of the neutrinosphere in a
self-consistent way (Gondek et al. 1997, hereafter G97). These
refinements enable us to study properties of PNS with arbitrary mass
(also with small mass, which consists essentially of subnuclear
density envelope alone) and find constraints on minimum and maximum
mass of NSs assuming conservation of the total baryon number of the
star during 3-30 seconds of its life (Bombaci 1996).

\section{Physical model of protoneutron star}

Our models of PNS are composed of a hot, neutrino-opaque interior,
separated from much colder, neutrino-transparent envelope by the
neutrinosphere. We consider two limiting thermal states of the hot
interior (see G97): isentropic, with entropy per baryon $s=1$ and
$s=2$; and isothermal, with $T_\infty=Te^{\nu(r)/2}=$const., where
$\nu (r)$ is the metric function (Zeldovich et al. 1971), $T_\infty$
and $T$ are the values of the temperature, measured by an observer at
infinity and by a local observer respectively. We chose $T=T_b=15 \ 
{\rm MeV}$ at the edge of the hot isothermal core.  The first case,
characteristic of a very initial state of a PNS corresponds to a
significant trapped lepton number $(Y_l=0.4)$. The second case
corresponds to situation after the deleptonization ($Y_\nu=0$) of a
PNS.  In both cases the EOS of hot dense matter is
determined using the moderately stiff model of
Lattimer and Swesty (1991)(LS). This is a standard model of dense
matter, composed of nucleons and leptons.  For high densities we
supplemented the LS model with contributions resulting from the
presence of neutrinos and antineutrinos of three flavours.

We consider PNS as an isolated, non-rotating, spherically
symmetric object, which has no magnetic field. Under these assumptions the
structure of the star is solved numerically using TOV equations
(Tolman 1939, Oppenheimer \& Volkoff 1939) for a given EOS. 
 Global parameters and static stability criteria
of PNS are discussed by G97.

\section{Results}
Static neutron stars with central density lower than central density
of the star with maximum mass $M_{\rm max}$, but greater than central
density of the star with minimum mass $M_{\rm min}$ are stable and can
exist (Harrison et al. 1965). In our case of moderately stiff equation
of state $M_{\rm max}({\rm NS})=2.044 M_\odot$ and $M_{\rm min}({\rm
NS})=0.054 M_\odot$. However, NS formed from PNS can have a
gravitational mass in significantly narrower region.  Since accretion
on the forming protoneutron star ceases $\sim 3$~s after birth
(Chevalier 1989) it is a good approximation to assume that during
transformation of PNS into NS (3-30 seconds after birth) the total
baryon number $A$ in the star is conserved. We compare NSs and PNSs by
fixing the total baryon (rest) mass of a star $M_{\rm bar}=Am_0$,
where $m_0$ is the mass of the hydrogen atom.

\subsubsection{Constraints on maximum mass of neutron stars}
In Fig. 1a, we plot gravitational mass as a function of the baryonic
mass for stable massive NSs and PNSs.  We see that the baryonic mass
$M_{\rm bar,max}$ corresponding to the maximum gravitational mass
$M_{\rm max}$ is the largest one for the static NSs (this is
consistent with results of Takatsuka 1995 and Bombaci 1996 for a
conventional equation of state).  For PNSs with isentropic core
(dashed and dotted-dashed lines) $M_{\rm bar,max}({\rm PNS})$ is
significantly lower than $M_{{\rm bar,max}}({\rm NS})$. In our case a
newborn isentropic PNS with the maximum gravitational mass
transforms, due to deleptonization and cooling, into a cold NS with
gravitational mass of 1.889-1.9 $M_\odot$ (points a, b at Fig. 1a).
NS with $M_{\rm {bar}}({\rm NS})>M_{\rm bar,max}({\rm
PNS})$ cannot be formed from a PNS.  The maximum mass of a NS
obtained from evolution is smaller than $M_{\rm max}({\rm NS})$
from neutron stars model by $\sim 0.15 M_\odot$ (for comparison
Takatsuka 1995 obtained 0.05-0.07 $M_\odot$).

\begin{figure}[t]
\begin{center}
\leavevmode
\epsfxsize=.45\columnwidth \epsfbox{ 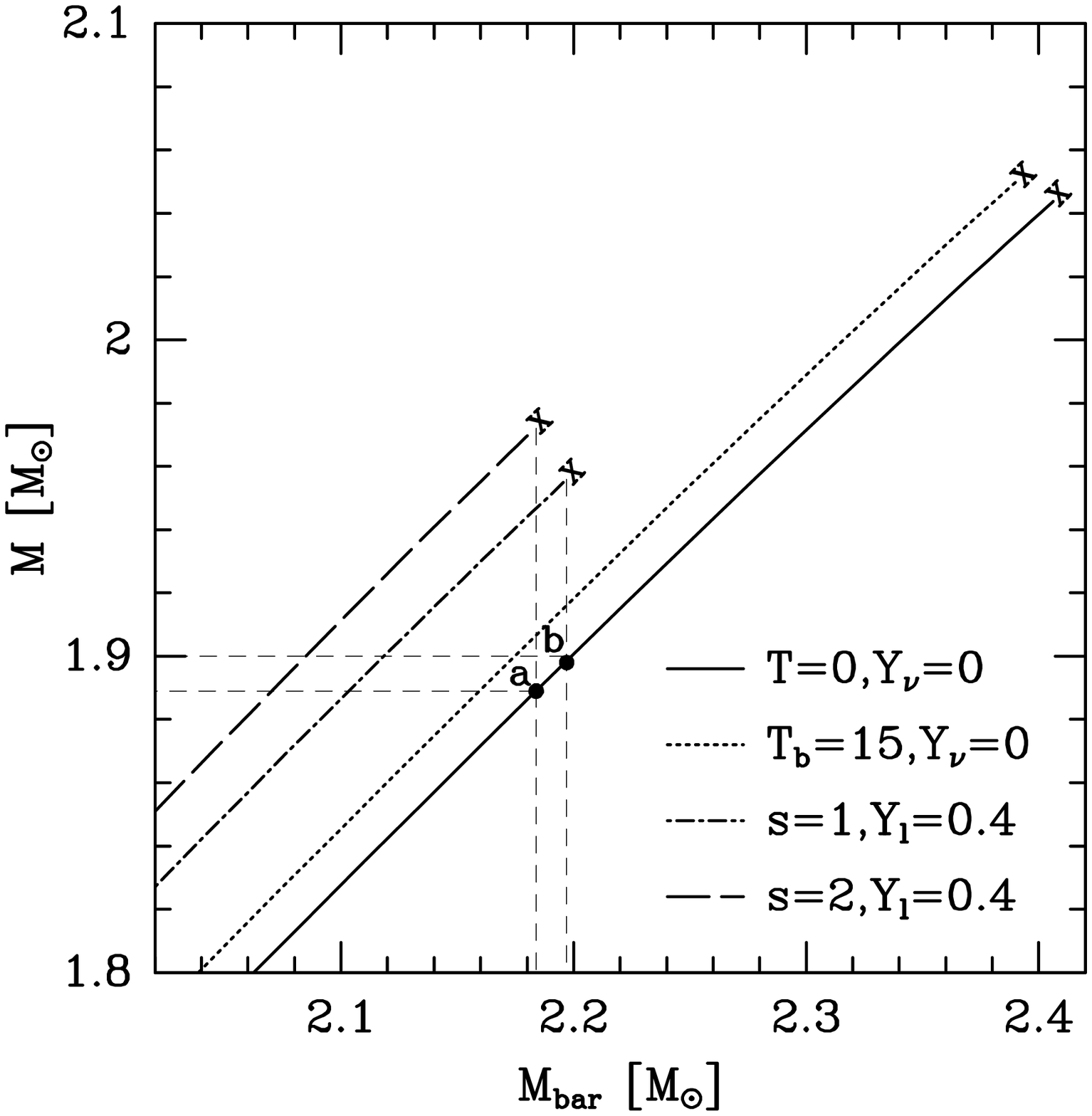}\hfil
\epsfxsize=.45\columnwidth \epsfbox{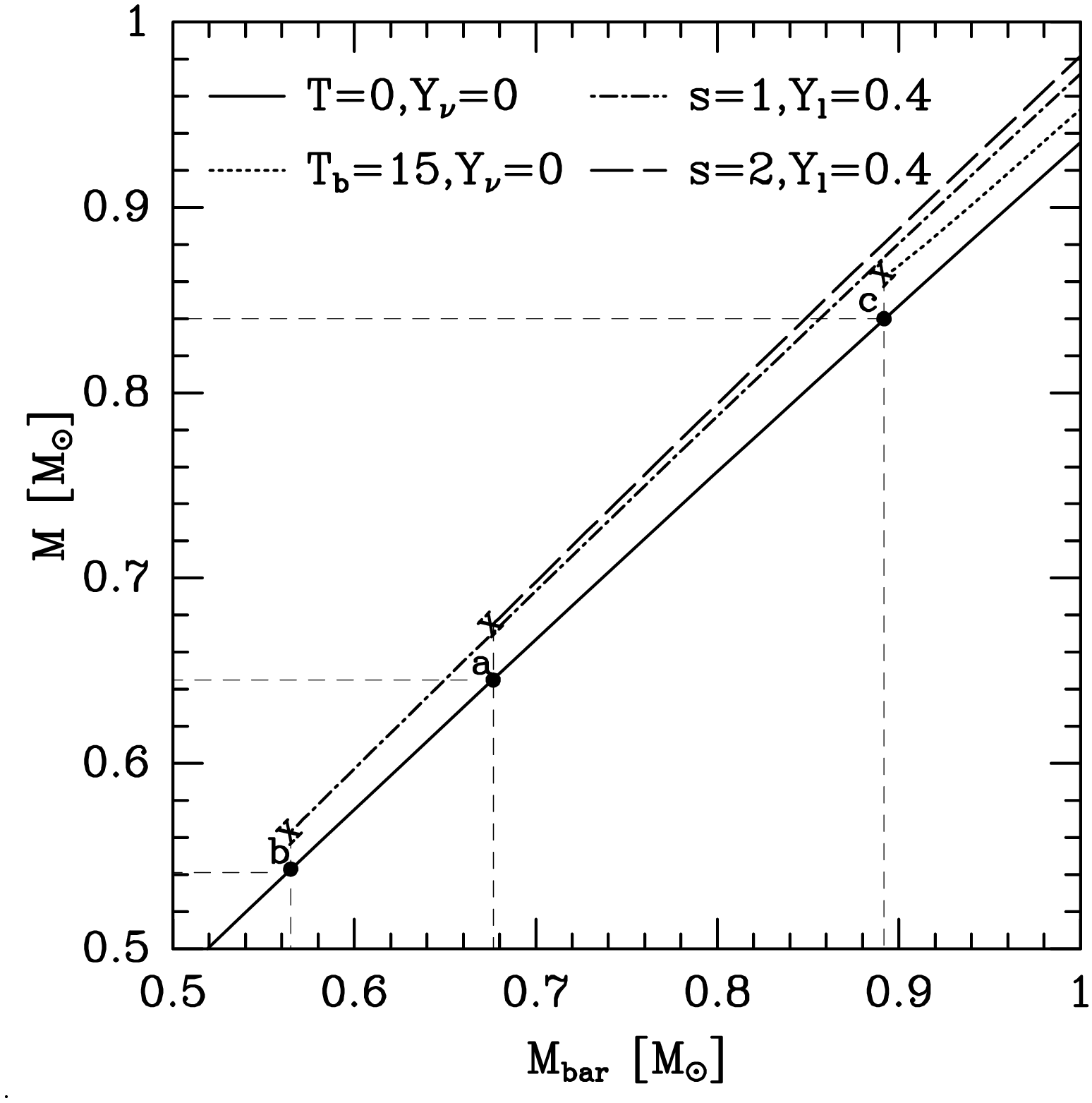}
\end{center}
\caption{a) Gravitational mass versus baryonic mass  for massive 
 protoneutron stars and neutron stars. The long dashed and dash-dotted
lines correspond to isentropic PNS stage ($t \sim 2-4$~s). The dotted
line represents isothermal PNS ($t \sim$ 10-20 s) and the solid line
corresponds to the final NSs. The crosses correspond to maximum
allowable masses for a given equation of state. The maximum baryon
mass for a PNS is smaller than the maximum allowed baryon mass for a
static NS. The points a and b represent NSs which has evolved from
isentropic PNSs with the maximum possible mass assuming
conservation of the total baryonic mass during evolution. 
b) Same as Fig. 1a, but for low massive protoneutron
stars and neutron stars. The crosses correspond to minimum allowable
masses of PNSs. The minimum baryon mass for a PNS is greater than the
minimum allowed baryon mass for a static NS ($M_{\rm bar,min}({\rm
NS})=0.055 M_\odot$). The points a, b, c represent NSs which has
evolved from PNSs with the minimum possible mass assuming conservation
of the total baryonic mass.  
} 
\label{mbduzo}
\end{figure}

\subsubsection{Constraints on minimum mass of neutron stars}

In Fig. 1b, we compare the $M$ versus $M_{\rm bar}$ between low mass
PNSs and NSs.  The baryonic mass $M_{\rm bar,min}$ corresponding to
the minimum gravitational mass $M_{\rm min}$ is the lowest one for static
neutron stars $M_{\rm bar,min}({\rm NS})=0.055 M_\odot$. We consider
two scenarios: an isentropic PNS transforms directly to a NS or
undergoes through a hot isothermal state.  In the first case the minimum
gravitational mass for NS is $0.54-0.65 M_\odot$ depending on the
value of entropy in isentropic core of a PNS, in the second case it is
$0.84 M_\odot $. Therefore NS with $M < 0.54 M_\odot$ ($M < 0.84
M_\odot$ if the PNS goes through the hot isothermal like configurations)
do not exist. 
\section{Conclusions}
We have constrained the minimum and maximum masses of NSs by the
considerations of their initial hot stage. All calculations were done
for a standard, moderately stiff dense matter equation of state.  We
show that the minimum gravitational mass of a NS is by about $0.5
M_\odot$ greater than $M_{\rm min}({\rm NS})$ in a static NS model,
while the maximum gravitational mass of a NS is by $\sim 0.15 M_\odot$
smaller than $M_{\rm max}({\rm NS}).$ The latter result seems to be
characteristic for a standard model of dense matter, composed of
nucleons and leptons (e.g. Takatsuka 1995, Bombaci 1996).
 The conclusions are different when we take into account
existence of an exotic components (hyperons, pions or kaons condensate
or quark matter) at hight densities (Bombaci et al. 1996). The
scenario of transformation of a PNS into a NS could be strongly
influenced by a phase transition in the central region of the
star. Appearance of exotic matter could dramatically soften the EOS of
dense matter, lowering maximum allowable mass of NSs.  In the case of
$K^-$ condensation, the maximum baryon mass of PNS is larger by $\sim
0.2~M_\odot$ than that of cold neutron stars (Brown \& Bethe 1994) and
the maximum mass of a NS is that one determined from a static NS
model. The deleptonization and cooling of PNSs of baryon mass
exceeding the maximum allowable baryon mass for NSs, would lead to
their collapse into BH.

\acknowledgments 
This work was supported in part by the following grants
KBN-2P03D01211, KBN-2P030D01413 and by the program R{\'e}seau
Formation Recherche of the French Minist{\`e}re de l'Enseignement
Sup{\'e}rieur et de la Recherche.

\end{document}